# Symmetric-Cyclic Bidirectional Quantum Teleportation of Bell-like State via Entanglement-Swapping

**Nikhita Singh[1*] and Ravi S Singh[1ƚ]**

[1] Department of Physics, Deen Dayal Upadhyaya Gorakhpur University, Gorakhpur (UP) - 273009, India.
E-mails: singhnikhitabhi6@gmail.com*, yesora27@gmail.com[ƚ]

## Abstract

Quantum teleportation is a primitive foundational protocol for exchanging quantum information in a quantum network as well as infrastructural operational strategy in the measurement-based quantum computation. Designing an efficient scheme for quantum teleportation is a vibrant field of intensive research. We propose a scheme wherein Bell-like states are being exchanged simultaneously in cyclic sequence, i.e., symmetric-cyclic bi-directional quantum teleportation, amongst three communicating parties forming a quantum network, Alice, Bob and Charlie via entanglement-swapping with the aid of a cluster of three maximally entangled GHZ-states as the quantum channel. Moreover, based upon communication- and operation- complexity, we compare our protocol with other equivalent protocols and found that the intrinsic efficiency of our protocol is maximum pegging at 33.33%.

---

## 1. Introduction

Quantum internet [1-3], being a large quantum network of networks [4,5] and a promising prospective quantum technology, may provide a common platform whereupon variant sorts of quantum information processing tasks such as quantum key distribution [6,7], quantum teleportation [8-11], quantum computation [12], quantum tele-computation [13], quantum secret sharing [14], quantum tele-cloning [15], quantum Broadcasting [16], quantum remote information concentration [17], quantum tele-amplification [18,19] etc., amongst nodes and internet-of-things devices [20,21] can be envisaged in unidirectional, bidirectional, cyclic sequence and multidirectional way following discrete-and continuous-variables quantum systems._Quantum networks furnish not only a basis of global quantum communication, but also, they are being deployed for distributed sensing [22,23], synchronization of clock [24] and blind quantum computation [25]. Intensive theoretical proposals have been put forth for robust and efficient networks utilizing novel techniques such as quantum repeaters [26-29], multiplexing [30,31] and advanced quantum state encodings [32,33]. Quite recently, quantum networks having variant constituent building blocks like quantum teleportation amongst non-neighbouring nodes [34], satellite to ground communication [35] and multiplexing [36,37] are experimentally demonstrated.

Quantum teleportation (QT) is the earliest primitive counterintuitive communication scheme in which the probability-amplitudes of an arbitrary quantum state possessed by a sender gets transplanted on the state of a quantum system preserved by a distantly located receiver. QT is theoretically discovered by Bennett et al. [8] in discrete-variable regime via exploiting nonclassical correlations of maximally entangled EPR-Bell pair [38,39]. QT is experimentally realized in physical systems such as photonic states [40], optical quantum modes [41], nuclear magnetic resonance [42], atomic ensembles [43,44], trapped atoms or solid-state systems [45,46] showcasing myriad quantum technological platforms.

In 1998, Karlsson and Bourennane proposed controlled quantum teleportation [47] via application of maximally entangled GHZ-state as the quantum channel, which may be regarded as first generalization of QT. Quite recently incorporating presence of adversary in communicating parties, controlled quantum teleportation has been employed to address the security issues [48]. QT protocol has outgrown in several strategy such as (symmetric and asymmetric) bi-directional quantum teleportation in discrete-and continuous-variable regime [49-52]. Researchers have adjoined controlled QT with bi-directional quantum teleportation protocols to propose many communications scheme in which multipartite quantum channels in noisy and noiseless environment have been deployed [53-55]. Multi-directional simultaneous quantum teleportation of arbitrary two quantum-bits between distant quantum nodes within a quantum network has been worked out via a nine-qubit entangled state as a quantum channel [56].

Cyclic QT, a unidirectional protocol for simultaneously exchanging quantum information encoded in single qubit in cyclic sequence, is proposed by Chen et al.[57] Myriad cyclic (controlled) QT (symmetric and asymmetric) [58-64] schemes has been proposed by employing multiparty quantum channels in discrete and continuous variable regimes. Notably, afore-described protocols are notably unidirectional having transmission of information to next participant either in clockwise or counter-clockwise direction. Quite recently, in 2019, a cyclic bidirectional QT via pseudo multi-qubit state is designed by Zhou et al. [65] wherein it is shown that amongst three legitimate users Alice, Bob and Charlie, possessing bi-qubit entangled states, Alice could transmit her quantum information to Bob and Charlie, respectively, Bob could transmit his quantum information to Alice and Charlie, respectively and at the same time, Charlie could transmit his qubits to Bob and Alice, respectively. Sun et al. [66] came up with a bi-directional cyclic controlled communication via a thirteen-qubits entangled state as the quantum channel, in which four parties are involved in communication strategy, each party could teleport their two different single-qubit states to the other two participants under supervision of a controller, respectively, in both clockwise and anticlockwise directions.

Zukowski et. al. [67] introduced a concept of entanglement swapping, experimentally validated by Pan et. al [68], for efficiently distributing entanglement amongst non-intertwined quantum systems. Notably, entanglement-swapping increases the distance of quantum communication and it, therefore, makes possible to realize long distance quantum communication in a quantum network [69-72]. A prime bottleneck in long distance quantum teleportation is that quantum information encoded in qubits over long distances deteriorates from losses and decoherence due to interaction with transmitting environment. Quantum repeater [26-29, 73] tracks these challenges by dividing the long distances into smaller segments and by applying entanglement swapping and classical communication to relay the quantum information across these segments. Wan et al. [72] investigated an entanglement swapping protocol using photon-number-encoded states that can effectively mitigate losses by tailoring the initial entangled states.

By leveraging entanglement swapping, a bidirectional QT scheme is put forth by Hassanpour et al. [50] wherein two parities (Alice and Bob) can transmit their unknown single qubit state via EPR-state serving as quantum channel. We, here, exploiting entanglement swapping technique, a novel symmetric-cyclic bidirectional QT scheme is strategized, in which Alice can communicate her Bell-like pair to Bob, Bob can transmit his Bell-like pair to Charlie, Charlie can send his Bell-like pair to Alice, and, in reverse direction, Alice can send her Bell-like pair to Charlie, Charlie can transmit his Bell-like pair to Bob, Bob can communicate his Bell-like pair to Alice, simultaneously. And finally, we have compared our protocol with previous protocols developed without entanglement swapping on the basis of classical and quantum resource consumption and found that our scheme witnessed notable increased intrinsic efficiency 33.33%.

The present investigation is structured in the following sections: preparation of quantum channel and a brief description of entanglement-swapping technique is presented in Section-2. In Section-3, we described our scheme in detail. Finally, In Section-4, we evaluated efficiency of our protocol, made a comparison with other existing protocols in literature and delineated future prospects pertaining to extension of the protocol in Section-5.

## 2. Preparation of Eighteen qubit entangled GHZ-state as Quantum Channel

First, we describe entanglement swapping technique before preparing the eighteen-qubit entangled state as a quantum channel for Cyclic bidirectional quantum teleportation. The tripartite GHZ-like state may be described by,

$$|G_0\rangle=\frac{|000\rangle+|111\rangle}{\sqrt{2}},|G_1\rangle=\frac{|000\rangle-|111\rangle}{\sqrt{2}},|G_2\rangle=\frac{|100\rangle+|011\rangle}{\sqrt{2}},$$

$$|G_3\rangle=\frac{|100\rangle-|011\rangle}{\sqrt{2}},|G_4\rangle=\frac{|010\rangle+|101\rangle}{\sqrt{2}},|G_5\rangle=\frac{|010\rangle-|101\rangle}{\sqrt{2}},\ |G_6\rangle=\frac{|110\rangle+|001\rangle}{\sqrt{2}},|G_7\rangle=\frac{|110\rangle-|001\rangle}{\sqrt{2}} \quad (1)$$

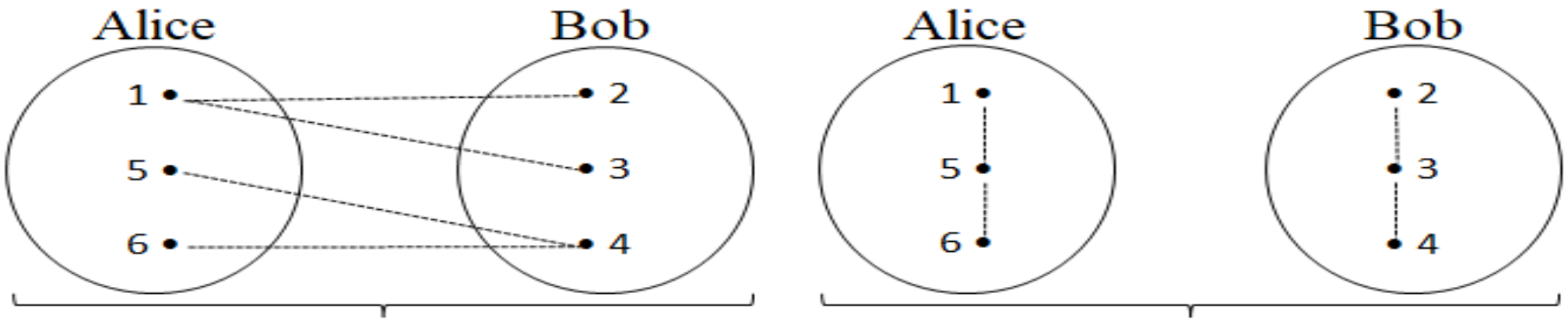

Before Entanglement swapping: Qubit 1, 2 &3 are mutually entangled and qubit 4, 5 & 6 are mutually entangled.

After Entanglement swapping: Qubit 1, 5 & 6 are mutually entangled and qubit 2, 3 & 4 are mutually entangled

Now, one may demonstrate entanglement-swapping process by taking one state out of the above tripartite GHZ-like states. We assume two parties Alice and Bob share tri-partite GHZ-state $|G_0\rangle_{123}$ and $|G_0\rangle_{456}$, wherein, qubits 1, 5 and 6 are with Alice and qubits 2, 3 and 4 are possessed by Bob. The global state of this system (Alice and Bob) may be given by,

$$\begin{aligned}
&|\psi\rangle_{123456}\\
&=|G_0\rangle_{123}\otimes|G_0\rangle_{456}\\
&=\left(\frac{|000\rangle+|111\rangle}{\sqrt{2}}\right)_{123}\otimes\left(\frac{|000\rangle+|111\rangle}{\sqrt{2}}\right)_{456}\\
&=\frac{1}{2}\left(|000\rangle_{123}|000\rangle_{456}+|000\rangle_{123}|111\rangle_{456}+|111\rangle_{123}|000\rangle_{456}+|111\rangle_{123}|111\rangle_{456}\right)\\
&=\frac{1}{2}\left(|000\rangle_{156}|000\rangle_{234}+|011\rangle_{156}|001\rangle_{234}+|100\rangle_{156}|110\rangle_{234}+|111\rangle_{156}|111\rangle_{234}\right)\\
&=\frac{1}{2}\left(|G_0\rangle_{156}|G_0\rangle_{234}+|G_1\rangle_{156}|G_1\rangle_{234}+|G_6\rangle_{156}|G_2\rangle_{234}+|G_7\rangle_{156}|G_3\rangle_{234}\right)
\end{aligned} \quad (2)$$

Now, if an observer performed GHZ-measurement by utilizing GHZ-basis, eq. (1), each result of measurement projects the global state, eq. (2), into one of the GHZ-like state, eq. (1), a maximally entangled tri-qubit state. Notably, if Alice and Bob share another state initially among GHZ-like state, eq. (1), similar result would appear by entangle-swapping technique.

A quantum channel for implementing our symmetric-cyclic bidirectional quantum teleportation, Sect. 3., is taken as a cluster of six GHZ states, eighteen-qubits entangled state,

$$|Q\rangle_{Channel}=|GHZ\rangle_{a_1b_2b_3}\otimes|GHZ\rangle_{a_2a_3b_1}\otimes|GHZ\rangle_{b_1'c_2c_3}\otimes|GHZ\rangle_{b_2'b_3'c_1}\otimes|GHZ\rangle_{c_1'a_2'a_3'}\otimes|GHZ\rangle_{c_2'c_3'a_1'} \quad (3)$$

Where, $|GHZ\rangle=\frac{(|000\rangle+|111\rangle)}{\sqrt{2}}=|G_0\rangle$

## 3. Symmetric-Cyclic bidirectional quantum teleportation protocol

We assume Alice, Bob and Charlie possess arbitrary unknown quantum information encoded in Bell-like states,

$$|\varphi\rangle_{A_1A_2} = \alpha_0|00\rangle + \alpha_1|11\rangle, |\varphi\rangle_{A_1'A_2'} = \beta_0|00\rangle + \beta_1|11\rangle,$$

$$|\varphi\rangle_{B_1B_2} = \mu_0|00\rangle + \mu_1|11\rangle, |\varphi\rangle_{B_1'B_2'} = \nu_0|00\rangle + \nu_1|11\rangle,$$

$$|\varphi\rangle_{C_1C_2} = \gamma_0|00\rangle + \gamma_1|11\rangle, |\varphi\rangle_{C_1'C_2'} = \lambda_0|00\rangle + \lambda_1|11\rangle, \tag{4}$$

respectively, where $\alpha_0$, $\alpha_1$, $\beta_0$, $\beta_1$, $\mu_0$, $\mu_1$, $\nu_0$, $\nu_1$, $\gamma_0$, $\gamma_1$, $\lambda_0$, $\lambda_1$ are unknown probability amplitudes and satisfy normalization conditions, $|\alpha_0|^2 + |\alpha_1|^2 = 1; |\beta_0|^2 + |\beta_1|^2 = 1;$ $|\mu_0|^2 + |\mu_1|^2 = 1; |\nu_0|^2 + |\nu_1|^2 = 1; |\gamma_0|^2 + |\gamma_1|^2 = 1; |\lambda_0|^2 + |\lambda_1|^2 = 1.$

Now, Alice strategizes to send her two Bell-like states, $|\phi\rangle_{A_1A_2}$, $|\phi\rangle_{A_1'A_2'}$, to Bob and Charlie, respectively, Bob sends, at the same instant, his two Bell-like states, $|\phi\rangle_{B_1B_2}$, $|\phi\rangle_{B_1'B_2'}$, to Alice and Charlie, respectively, and Charlie transmits, simultaneously, his two Bell-like states, $|\phi\rangle_{C_1C_2}$, $|\phi\rangle_{C_1'C_2'}$, to Bob and Alice respectively. All afore-mentioned transmission of quantum information, i.e., symmetric-cyclic bidirectional quantum teleportation, simultaneously, can be implemented among three communicating parties (Alice, Bob and Charlie) with sharing of an eighteen-qubits entangled state employed as the quantum channel, Eq.3, of which qubits are distributed among communicating parties in specific orders: $(a_1, a_2, a_3, a_1', a_2'. a_3')$ are possessed by Alice, $(b_1, b_2, b_3, b_1', b_2'. b_3')$ are with Bob, $(c_1, c_2, c_3, c_1', c_2'. c_3')$ belongs to Charlie (see Fig.1). The global state of entire system would be,

$$\begin{aligned}
&|\zeta\rangle_{Whole} \\
&= |\varphi\rangle_{A_1A_2} \otimes |\varphi\rangle_{B_1B_2} \otimes |\varphi\rangle_{B_1'B_2'} \otimes |\varphi\rangle_{C_1C_2} \otimes |\varphi\rangle_{C_1'C_2'} \otimes |\varphi\rangle_{A_1'A_2'} \otimes |Q\rangle_{Channel} \\
&= \left(\alpha_0|00\rangle + \alpha_1|11\rangle\right)_{A_1A_2} \otimes \left(\mu_0|00\rangle + \mu_1|11\rangle\right)_{B_1B_2} \otimes \left(\nu_0|00\rangle + \nu_1|11\rangle\right)_{B_1'B_2'} \otimes \left(\gamma_0|00\rangle + \gamma_1|11\rangle\right)_{C_1C_2} \otimes \left(\lambda_0|00\rangle + \lambda_1|11\rangle\right)_{C_1'C_2'} \\
&\otimes \left(\beta_0|00\rangle + \beta_1|11\rangle\right)_{A_1'A_2'} \otimes |GHZ\rangle_{a_1b_2b_3} \otimes |GHZ\rangle_{a_2a_3b_1} \otimes |GHZ\rangle_{b_1'c_2c_3} \otimes |GHZ\rangle_{b_2'b_3'c_1} \otimes |GHZ\rangle_{c_1'a_2'a_3'} \otimes |GHZ\rangle_{c_2'c_3'a_1'}
\end{aligned} \tag{5}$$

where equations (3) and (4) have been used. Inserting $|GHZ\rangle = \left(|000\rangle + |111\rangle\right)/\sqrt{2}$ in equation (5), one obtains,

$$\begin{aligned}
&|\zeta\rangle_{Whole} \\
&= \left(\alpha_0|00\rangle + \alpha_1|11\rangle\right)_{A_1A_2} \otimes \left(\mu_0|00\rangle + \mu_1|11\rangle\right)_{B_1B_2} \otimes \left(\nu_0|00\rangle + \nu_1|11\rangle\right)_{B_1'B_2'} \\
&\otimes \left(\gamma_0|00\rangle + \gamma_1|11\rangle\right)_{C_1C_2} \otimes \left(\lambda_0|00\rangle + \lambda_1|11\rangle\right)_{C_1'C_2'} \otimes \left(\beta_0|00\rangle + \beta_1|11\rangle\right)_{A_1'A_2'} \otimes \frac{1}{8}\{ \left(|000\rangle + |111\rangle\right)_{a_1b_2b_3} \otimes \left(|000\rangle + |111\rangle\right)_{a_2a_3b_1} \\
&\otimes \left(|000\rangle + |111\rangle\right)_{b_1'c_2c_3} \otimes \left(|000\rangle + |111\rangle\right)_{b_2'b_3'c_1} \otimes \left(|000\rangle + |111\rangle\right)_{c_1'a_2'a_3'} \otimes \left(|000\rangle + |111\rangle\right)_{c_2'c_3'a_1'} \}
\end{aligned} \tag{6}$$

To accomplish symmetric-cyclic bidirectional quantum teleportation scheme, following steps are involved-

***Step-1*** Alice, Bob and Charlie apply Controlled-NOT operations on qubit-pairs$(A_1, a_1)$, $(B_1, b_1)$, $(B'_1, b'_1)$, $(C_1, c_1)$, $(C'_1, c'_1)$, $(A'_1, a'_1)$ with first qubit as control qubit and second as a target qubit, respectively. Consequently the state, (Eq.6), gets transformed into,

$$
\begin{aligned}
|\zeta'\rangle_{Whole} = \frac{1}{8}\Big\{ & \big(|000000\rangle + |000111\rangle + |111000\rangle + |111111\rangle\big)_{a_1b_2b_3a_2a_3b_1} \alpha_0\mu_0 |0000\rangle_{A_1A_2B_1B_2} \\
& + \big(|000001\rangle + |000110\rangle + |111001\rangle + |111110\rangle\big)_{a_1b_2b_3a_2a_3b_1} \alpha_0\mu_1 |0011\rangle_{A_1A_2B_1B_2} \\
& + \big(|100000\rangle + |100111\rangle + |011000\rangle + |011111\rangle\big)_{a_1b_2b_3a_2a_3b_1} \alpha_1\mu_0 |1100\rangle_{A_1A_2B_1B_2} \\
& + \big(|100001\rangle + |100110\rangle + |011001\rangle + |011110\rangle\big)_{a_1b_2b_3a_2a_3b_1} \alpha_1\mu_1 |1111\rangle_{A_1A_2B_1B_2} \Big\} \\
\otimes \Big\{ & \big(|000000\rangle + |000111\rangle + |111000\rangle + |111111\rangle\big)_{b'_1c_2c_3b'_2b'_3c_1} \nu_0\gamma_0 |0000\rangle_{B'_1B'_2C_1C_2} \\
& + \big(|000001\rangle + |000110\rangle + |111001\rangle + |111110\rangle\big)_{b'_1c_2c_3b'_2b'_3c_1} \nu_0\gamma_1 |0011\rangle_{B'_1B'_2C_1C_2} \\
& + \big(|100000\rangle + |100111\rangle + |011000\rangle + |011111\rangle\big)_{b'_1c_2c_3b'_2b'_3c_1} \nu_1\gamma_0 |1100\rangle_{B'_1B'_2C_1C_2} \\
& + \big(|100001\rangle + |100110\rangle + |011001\rangle + |011110\rangle\big)_{b'_1c_2c_3b'_2b'_3c_1} \nu_1\gamma_1 |1111\rangle_{B'_1B'_2C_1C_2} \Big\} \\
\otimes \Big\{ & \big(|000000\rangle + |000111\rangle + |111000\rangle + |111111\rangle\big)_{c'_1a'_2a'_3c'_2c'_3a'_1} \lambda_0\beta_0 |0000\rangle_{C'_1C'_2A'_1A'_2} \\
& + \big(|000001\rangle + |000110\rangle + |111001\rangle + |111110\rangle\big)_{c'_1a'_2a'_3c'_2c'_3a'_1} \lambda_0\beta_1 |0011\rangle_{C'_1C'_2A'_1A'_2} \\
& + \big(|100000\rangle + |100111\rangle + |011000\rangle + |011111\rangle\big)_{c'_1a'_2a'_3c'_2c'_3a'_1} \lambda_1\beta_0 |1100\rangle_{C'_1C'_2A'_1A'_2} \\
& + \big(|100001\rangle + |100110\rangle + |011001\rangle + |011110\rangle\big)_{c'_1a'_2a'_3c'_2c'_3a'_1} \lambda_1\beta_1 |1111\rangle_{C'_1C'_2A'_1A'_2} \Big\}
\end{aligned} \tag{7}
$$

Applying entanglement-swapping eq. (7), using eq (1), yields,

$$
\begin{aligned}
|\zeta'\rangle_{Whole} = \frac{1}{8}\Big[ & \Big\{ |G_0\rangle_{a_1a_2a_3}|G_0\rangle_{b_1b_2b_3} + |G_1\rangle_{a_1a_2a_3}|G_1\rangle_{b_1b_2b_3} + |G_2\rangle_{a_1a_2a_3}|G_6\rangle_{b_1b_2b_3} + |G_3\rangle_{a_1a_2a_3}|G_7\rangle_{b_1b_2b_3} \Big\} \alpha_0\mu_0 |0000\rangle_{A_1A_2B_1B_2} \\
& + \Big\{ |G_0\rangle_{a_1a_2a_3}|G_6\rangle_{b_1b_2b_3} - |G_1\rangle_{a_1a_2a_3}|G_7\rangle_{b_1b_2b_3} + |G_2\rangle_{a_1a_2a_3}|G_0\rangle_{b_1b_2b_3} - |G_3\rangle_{a_1a_2a_3}|G_1\rangle_{b_1b_2b_3} \Big\} \alpha_0\mu_1 |0011\rangle_{A_1A_2B_1B_2} \\
& + \Big\{ |G_2\rangle_{a_1a_2a_3}|G_0\rangle_{b_1b_2b_3} + |G_3\rangle_{a_1a_2a_3}|G_1\rangle_{b_1b_2b_3} + |G_0\rangle_{a_1a_2a_3}|G_6\rangle_{b_1b_2b_3} + |G_1\rangle_{a_1a_2a_3}|G_7\rangle_{b_1b_2b_3} \Big\} \alpha_1\mu_0 |1100\rangle_{A_1A_2B_1B_2} \\
& + \Big\{ |G_2\rangle_{a_1a_2a_3}|G_6\rangle_{b_1b_2b_3} - |G_3\rangle_{a_1a_2a_3}|G_7\rangle_{b_1b_2b_3} + |G_0\rangle_{a_1a_2a_3}|G_0\rangle_{b_1b_2b_3} - |G_1\rangle_{a_1a_2a_3}|G_1\rangle_{b_1b_2b_3} \Big\} \alpha_1\mu_1 |1111\rangle_{A_1A_2B_1B_2} \Big] \\
\otimes \Big[ & \Big\{ |G_0\rangle_{b'_1b'_2b'_3}|G_0\rangle_{c_1c_2c_3} + |G_1\rangle_{b'_1b'_2b'_3}|G_1\rangle_{c_1c_2c_3} + |G_2\rangle_{b'_1b'_2b'_3}|G_6\rangle_{c_1c_2c_3} + |G_3\rangle_{b'_1b'_2b'_3}|G_7\rangle_{c_1c_2c_3} \Big\} \nu_0\gamma_0 |0000\rangle_{B'_1B'_2C_1C_2} \\
& + \Big\{ |G_0\rangle_{b'_1b'_2b'_3}|G_6\rangle_{c_1c_2c_3} - |G_1\rangle_{b'_1b'_2b'_3}|G_7\rangle_{c_1c_2c_3} + |G_2\rangle_{b'_1b'_2b'_3}|G_0\rangle_{c_1c_2c_3} - |G_3\rangle_{b'_1b'_2b'_3}|G_1\rangle_{c_1c_2c_3} \Big\} \nu_0\gamma_1 |0011\rangle_{B'_1B'_2C_1C_2} \\
& + \Big\{ |G_2\rangle_{b'_1b'_2b'_3}|G_0\rangle_{c_1c_2c_3} + |G_3\rangle_{b'_1b'_2b'_3}|G_1\rangle_{c_1c_2c_3} + |G_0\rangle_{b'_1b'_2b'_3}|G_6\rangle_{c_1c_2c_3} + |G_1\rangle_{b'_1b'_2b'_3}|G_7\rangle_{c_1c_2c_3} \Big\} \nu_1\gamma_0 |1100\rangle_{B'_1B'_2C_1C_2} \\
& + \Big\{ |G_2\rangle_{b'_1b'_2b'_3}|G_6\rangle_{c_1c_2c_3} - |G_3\rangle_{b'_1b'_2b'_3}|G_7\rangle_{c_1c_2c_3} + |G_0\rangle_{b'_1b'_2b'_3}|G_0\rangle_{c_1c_2c_3} - |G_1\rangle_{b'_1b'_2b'_3}|G_1\rangle_{c_1c_2c_3} \Big\} \nu_1\gamma_1 |1111\rangle_{B'_1B'_2C_1C_2} \Big] \\
\otimes \Big[ & \Big\{ |G_0\rangle_{c'_1c'_2c'_3}|G_0\rangle_{a'_1a'_2a'_3} + |G_1\rangle_{c'_1c'_2c'_3}|G_1\rangle_{a'_1a'_2a'_3} + |G_2\rangle_{c'_1c'_2c'_3}|G_6\rangle_{a'_1a'_2a'_3} + |G_3\rangle_{c'_1c'_2c'_3}|G_7\rangle_{a'_1a'_2a'_3} \Big\} \lambda_0\beta_0 |0000\rangle_{C'_1C'_2A'_1A'_2} \\
& + \Big\{ |G_0\rangle_{c'_1c'_2c'_3}|G_6\rangle_{a'_1a'_2a'_3} - |G_1\rangle_{c'_1c'_2c'_3}|G_7\rangle_{a'_1a'_2a'_3} + |G_2\rangle_{c'_1c'_2c'_3}|G_0\rangle_{a'_1a'_2a'_3} - |G_3\rangle_{c'_1c'_2c'_3}|G_1\rangle_{a'_1a'_2a'_3} \Big\} \lambda_0\beta_1 |0000\rangle_{C'_1C'_2A'_1A'_2} \\
& + \Big\{ |G_2\rangle_{c'_1c'_2c'_3}|G_0\rangle_{a'_1a'_2a'_3} + |G_3\rangle_{c'_1c'_2c'_3}|G_1\rangle_{a'_1a'_2a'_3} + |G_0\rangle_{c'_1c'_2c'_3}|G_6\rangle_{a'_1a'_2a'_3} + |G_1\rangle_{c'_1c'_2c'_3}|G_7\rangle_{a'_1a'_2a'_3} \Big\} \lambda_1\beta_0 |0000\rangle_{C'_1C'_2A'_1A'_2} \\
& + \Big\{ |G_2\rangle_{c'_1c'_2c'_3}|G_6\rangle_{a'_1a'_2a'_3} - |G_3\rangle_{c'_1c'_2c'_3}|G_7\rangle_{a'_1a'_2a'_3} + |G_0\rangle_{c'_1c'_2c'_3}|G_0\rangle_{a'_1a'_2a'_3} - |G_1\rangle_{c'_1c'_2c'_3}|G_1\rangle_{a'_1a'_2a'_3} \Big\} \lambda_1\beta_1 |0000\rangle_{C'_1C'_2A'_1A'_2} \Big]
\end{aligned}
$$

***Step-2*** Alice, Bob and Charlie perform simultaneous single-qubit measurement using Z-basis, $\{|0\rangle,|1\rangle\}$, on qubits $a_1$, $b_1$, $b_1'$, $c_1$, $c_1'$, $a_1'$, and X-basis measurement using basis, $\{|+\rangle,|-\rangle\}$ on qubits $A_1$, $B_1$, $B_1'$, $C_1$, $C_1'$, $A_1'$, respectively. Next, they communicate their results of measurements to each other via classical channel.

As an example, if Alice's measurement-results are $|+\rangle_{A_1}\otimes|0\rangle_{a_1},|+\rangle_{A_1'}\otimes|0\rangle_{a_1'}$, Bob's measurement results are $|+\rangle_{B_1}\otimes|0\rangle_{b_1},|+\rangle_{B_1'}\otimes|0\rangle_{b_1'}$ and Charlie's measurement-results are $|+\rangle_{C_1}\otimes|0\rangle_{c_1},|+\rangle_{C_1'}\otimes|0\rangle_{c_1'}$, then global state would collapse to,

$$\begin{aligned}
&\left({}_{a_1'}\langle 0|\otimes{}_{A_1'}\langle +|\right)\otimes\left({}_{c_1'}\langle 0|\otimes{}_{C_1'}\langle +|\right)\otimes\left({}_{c_1}\langle 0|\otimes{}_{C_1}\langle +|\right)\otimes\left({}_{b_1'}\langle 0|\otimes{}_{B_1'}\langle +|\right)\otimes\left({}_{b_1}\langle 0|\otimes{}_{B_1}\langle +|\right)\otimes\left({}_{a_1}\langle 0|\otimes{}_{A_1}\langle +|\right)|\zeta'\rangle_{Whole}\\
&=|\zeta''\rangle_{b_2b_3a_2a_3A_2B_2c_2c_3b_2'b'B_2'C_2a_2'a_3'c_2'c_3'C_2'A_2'}\\
&=\left\{\left(\alpha_0\mu_0|000000\rangle+\alpha_0\mu_1|001101\rangle+\alpha_1\mu_0|110010\rangle+\alpha_1\mu_1|111111\rangle\right)_{b_2b_3a_2a_3A_2B_2}\right\}\\
&\otimes\left\{\left(\nu_0\gamma_0|000000\rangle+\nu_0\gamma_1|001101\rangle+\nu_1\gamma_0|110010\rangle+\nu_1\gamma_1|111111\rangle\right)_{c_2c_3b_2'b_3'B_2'C_2}\right\}\\
&\otimes\left\{\left(\lambda_0\beta_0|000000\rangle+\lambda_0\beta_1|001101\rangle+\lambda_1\beta_0|110010\rangle+\lambda_1\beta_1|111111\rangle\right)_{a_2'a_3'c_2'c_3'C_2'A_2'}\right\}
\end{aligned}\tag{8}$$

***Step-3*** Again, Alice, Bob and Charlie perform X-basis measurement, $\{|+\rangle,|-\rangle\}$ on qubits $A_2$, $B_2$, $B_2'$, $C_2$, $C_2'$, $A_2'$, respectively and their measurement-results are communicated to each other via classical channel. Here 64-different measurement-results will be appearing. Some of these measurement-results along with their unitary operations are demonstrated in Table-1(see appendix-A). For example, if Alice's measurement-results are $|+\rangle_{A_2},|+\rangle_{A_2'}$, Bob's measurement results are $|+\rangle_{B_2},|+\rangle_{B_2'}$ and Charlie's measurement results are $|+\rangle_{C_2},|+\rangle_{C_2'}$, the global state would collapse to,

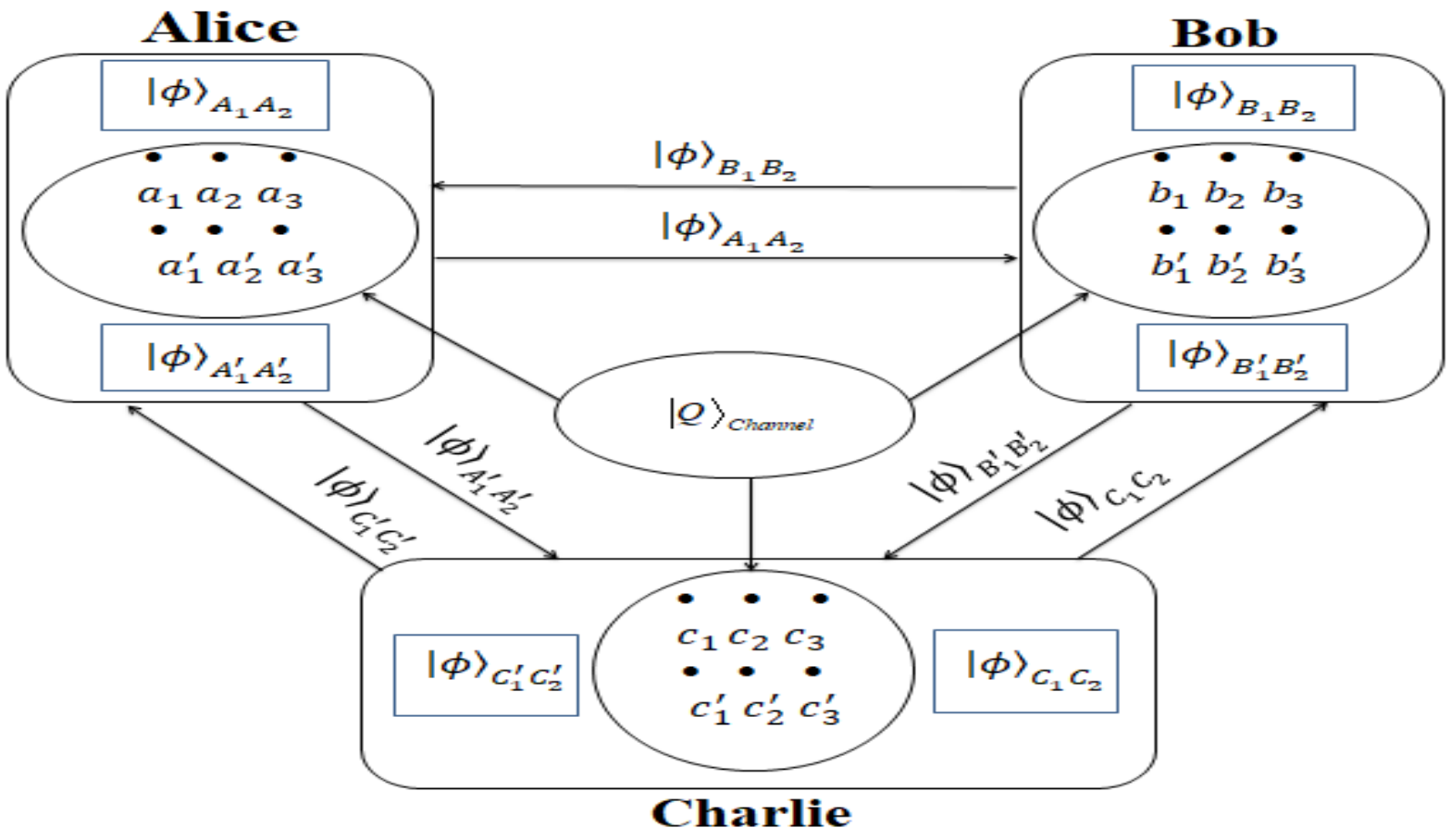

**Figure.1** Schematic diagram of symmetric-cyclic bidirectional quantum teleportation protocol. Here, $|\varphi\rangle_{A_1A_2}$ and $|\varphi\rangle_{A_1'A_2'}$ are Alice's information state, $|\varphi\rangle_{B_1B_2}$ and $|\varphi\rangle_{B_1'B_2'}$ are Bob's information state, $|\varphi\rangle_{C_1C_2}$ and $|\varphi\rangle_{C_1'C_2'}$ are Charlie's information state. Qubits $a_1$, $a_2$, $a_3$, $a_1'$, $a_2'$, $a_3'$, $b_1$, $b_2$, $b_3$, $b_1'$, $b_2'$, $b_3'$, $c_1$, $c_2$, $c_3$, $c_1'$, $c_2'$, $c_3'$ used in formation of quantum channel. A detailed method of this protocol is given in Sect. 3.

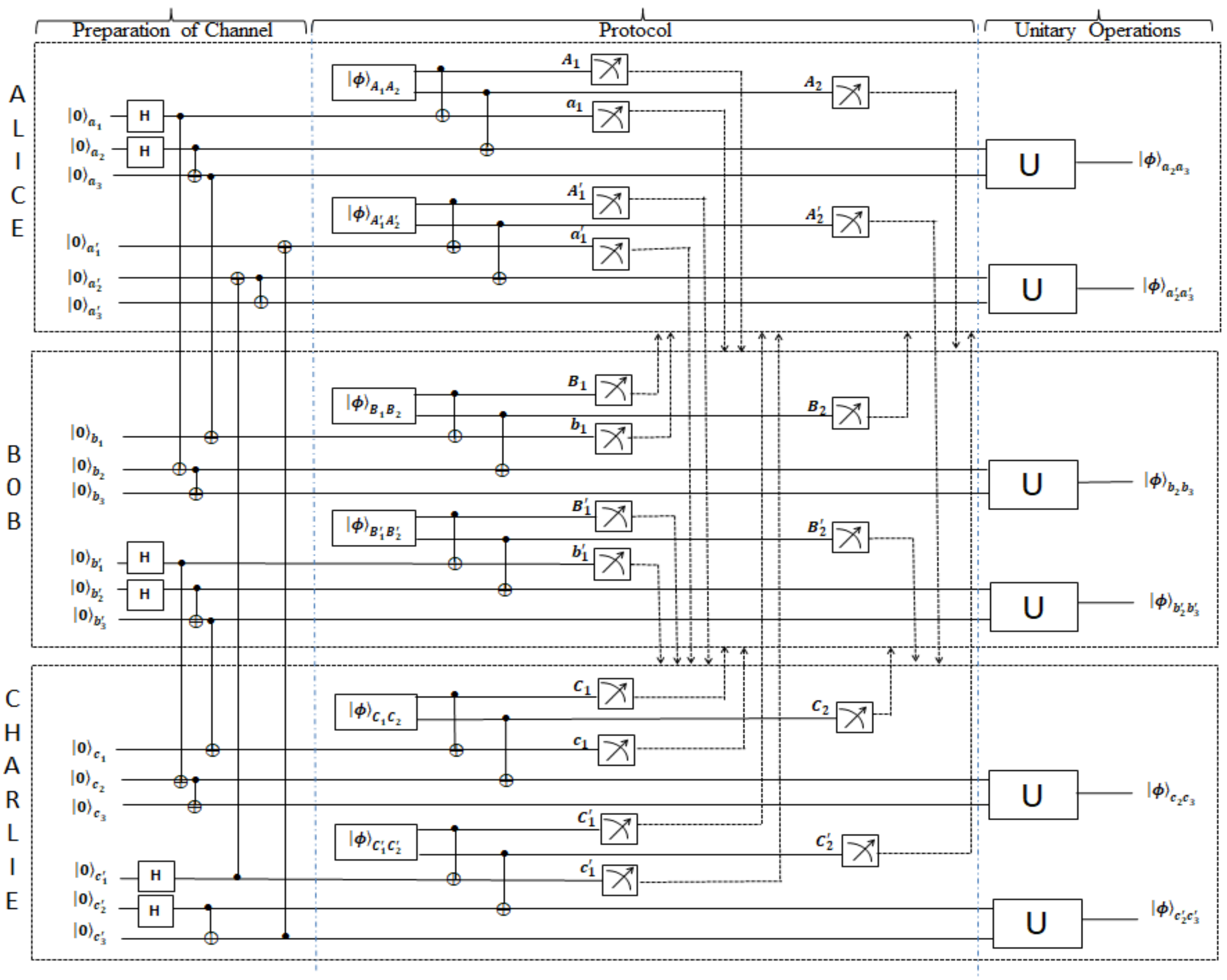

**Figure.2** Quantum circuit for symmetric-cyclic bidirectional Quantum Teleportation protocol of Bell pairs among three parties Alice, Bob and Charlie via GHZ-state. Here H represents Hadamard-gate operations. Here, X-basis measurements has been done on qubits $A_1$, $B_1$, $B_1'$, $C_1$, $C_1'$, $A_1'$, $A_2$, $B_2$, $B_2'$, $C_2$, $C_2'$, $A_2'$ and Z-basis measurements on qubits $a_1$, $b_1$, $b_1'$, $c_1$, $c_1'$, $a_1'$. **U** represents unitary operations. Dashed lines represent classical communications.

$$
\begin{aligned}
&\left({}_{A_2'}\langle +|\otimes {}_{C_2'}\langle +|\right)\otimes\left({}_{C_2}\langle +|\otimes {}_{B_2'}\langle +|\right)\otimes\left({}_{B_2}\langle +|\otimes {}_{A_2}\langle +|\right)|\zeta''\rangle_{b_2b_3a_2a_3A_2B_2c_2c_3b_2'b'B_2'C_2a_2'a_3'c_2'c_3'C_2'A_2'}\\
&=|\zeta'''\rangle_{b_2b_3a_2a_3c_2c_3b_2'b'a_2'a_3'c_2'c_3'}\\
&=\left\{\left(\alpha_0\mu_0|0000\rangle+\alpha_0\mu_1|0011\rangle+\alpha_1\mu_0|1100\rangle+\alpha_1\mu_1|1111\rangle\right)_{b_2b_3a_2a_3}\right\}\\
&\otimes\left\{\left(\nu_0\gamma_0|0000\rangle+\nu_0\gamma_1|0011\rangle+\nu_1\gamma_0|1100\rangle+\nu_1\gamma_1|1111\rangle\right)_{c_2c_3b_2'b_3'}\right\}\\
&\otimes\left\{\left(\lambda_0\beta_0|0000\rangle+\lambda_0\beta_1|0011\rangle+\lambda_1\beta_0|1100\rangle+\lambda_1\beta_1|1111\rangle\right)_{a_2'a_3'c_2'c_3'}\right\}
\end{aligned}
\tag{9}
$$

Rewriting one obtains,

$$
\begin{aligned}
&|\zeta'''\rangle_{b_2b_3a_2a_3c_2c_3b_2'b'a_2'a_3'c_2'c_3'}\\
&=\left(\alpha_0|00\rangle+\alpha_1|11\rangle\right)_{b_2b_3}\otimes\left(\mu_0|00\rangle+\mu_1|11\rangle\right)_{a_2a_3}\otimes\left(\nu_0|00\rangle+\nu_1|11\rangle\right)_{c_2c_3}\otimes\left(\gamma_0|00\rangle+\gamma_1|11\rangle\right)_{b_2'b_3'}\otimes\left(\lambda_0|00\rangle+\lambda_1|11\rangle\right)_{a_2'a_3'}\otimes\left(\beta_0|00\rangle+\beta_1|11\rangle\right)_{c_2'c_3'}
\end{aligned}
$$

Step 4: Finally, depending upon measurement result received from sender (see eq.9) Alice, Bob and Charlie will apply appropriate unitary operations as, $\hat{I}_{a_2} \otimes \hat{I}_{a_3} \otimes \hat{I}_{a'_2} \otimes \hat{I}_{a'_3}$ , $\hat{I}_{b_2} \otimes \hat{I}_{b_3} \otimes \hat{I}_{b'_2} \otimes \hat{I}_{b'_3}$ & $\hat{I}_{c_2} \otimes \hat{I}_{c_3} \otimes \hat{I}_{c'_2} \otimes \hat{I}_{c'_3}$ respectively, to reconstruct the desired information states.

That is to say, following afore-described steps Alice, Bob and Charlie transmitted their Bell-like pairs cyclically in both clockwise and anticlockwise directions. Hence, symmetric-cyclic bidirectional quantum teleportation Protocol can be realized faithfully. One must recognize that total $2^{18}$ different measurement-results may appear which have equal probability $1/2^{18}$. Since, here, it is not possible to show all these possible measurement-results, we have shown only a few measurement-results of Step.3 in Appendix-A, table-2.

| **Protocols** | **Initial State (Alice, Bob, Charlie )** | **Q-C** | **QIT** | **C-bits** | **Entanglement Swapping** | **Efficiency ($\eta$)** |
|---|---|---|---|---|---|---|
| Y. Li et al. | 2-2-2 | 10 | 6 | 9 | No | 31.5% |
| Z. W. Sang | 1-1-1 | 7 | 3 | 7 | No | 21.42% |
| R. Rahmawati et al. | 1-1-1<br>2-2-2<br>1-2-3 | 9<br>12<br>11 | 3<br>6<br>6 | 9<br>9<br>9 | No<br>No<br>No | 16.7%<br>28.57%<br>30% |
| Our | 2-2-2<br>2-2-2 | 18 | 12 | 18 | Yes | 33.33% |

**Table.1** Comparison between contemporary schemes for cyclic QT. Here, QC represents the total number of qubits employed in quantum channel, QIT represents the total number of qubits teleported, and C-bits represents the total number of classical bits consumed in classical communications.

## 4. Efficiency Analysis

A figure of merit in QT, namely, intrinsic efficiency, is introduced by Yuan et al. [72] as $\eta = q_s/(q_u + b_t)$, where $q_s$ is the number of qubits in the 'to be transmitted' information-state, $q_u$ is the number of qubits in quantum channel and $b_t$ is number of classical bits used in scheme. Applying this formula, efficiency of our protocol pegs at 33.33 %. Further, we compare recently-designed cyclic quantum teleportation protocols, wherein entanglement swapping is not invoked, on the basis of classical and quantum resource consumption (see Table-2).

## 5. Conclusion

We designed, using entanglement swapping technique, a scheme of symmetric cyclic bidirectional QT via a cluster of GHZ-states as a quantum channel wherein Alice, Bob and Charlie act as senders as well as receivers. We have utilized, in our protocol, only X-basis and Z- basis (computational basis) measurements, easily implementable in laboratory as compared to Bell-basis measurement. Since we have invoked entanglement swapping technique into the protocol, the scheme may be translated for large distance quantum communication.

## Acknowledgements

We acknowledge Prof. Lev Vaidman, School of Physics and Astronomy, Tel Aviv University, Israel and Prof. Mark M. Wilde, School of Electrical and Computer Engineering Cornell University, New York for drawing our attention toward their investigations on continuous variable quantum teleportation and bidirectional quantum teleportation. One of the authors (RSS) is also highly indebted for the discussions with Prof. R. R. Puri, Indian Institute of Technology, Gandhinagar (Gujarat), India.

**Appendix. A**

**Table.2** Measurements results and Unitary operations performed by Alice, Bob and Charlie. Here, $\hat{I}$ represents identity operator which is defined as $\hat{I} = |0\rangle\langle 0| + |1\rangle\langle 1|$ and other three (Pauli) unitary operators are defined as $\hat{\sigma}_x = |0\rangle\langle 1| + |1\rangle\langle 0|, \hat{\sigma}_y = i(|0\rangle\langle 1| - |1\rangle\langle 0|), \hat{\sigma}_z = |0\rangle\langle 0| - |1\rangle\langle 1|$.

| Alice`s Measurement Result | Bob`s Measurement Result | Charlie's Measurement Result | Alice`s Unitary Operation | Bob`s Unitary Operation | Charlie's Unitary Operation |
|---|---|---|---|---|---|
| $\lvert+\rangle_{A_2}\lvert+\rangle_{A_2'}$ | $\lvert+\rangle_{B_2}\lvert+\rangle_{B_2'}$ | $\lvert+\rangle_{C_2}\lvert+\rangle_{C_2'}$ | $\hat{I}_{a_2}\otimes\hat{I}_{a_3}\otimes\hat{I}_{a_2'}\otimes\hat{I}_{a_3'}$ | $\hat{I}_{b_2}\otimes\hat{I}_{b_3}\otimes\hat{I}_{b_2'}\otimes\hat{I}_{b_3'}$ | $\hat{I}_{c_2}\otimes\hat{I}_{c_3}\otimes\hat{I}_{c_2'}\otimes\hat{I}_{c_3'}$ |
| $\lvert+\rangle_{A_2}\lvert-\rangle_{A_2'}$ | $\lvert+\rangle_{B_2}\lvert+\rangle_{B_2'}$ | $\lvert+\rangle_{C_2}\lvert+\rangle_{C_2'}$ | $\hat{I}_{a_2}\otimes\hat{I}_{a_3}\otimes\hat{I}_{a_2'}\otimes\hat{I}_{a_3'}$ | $\hat{I}_{b_2}\otimes\hat{I}_{b_3}\otimes\hat{I}_{b_2'}\otimes\hat{I}_{b_3'}$ | $\hat{I}_{c_2}\otimes\hat{I}_{c_3}\otimes\hat{I}_{c_2'}\otimes\hat{\sigma}^z_{c_3'}$ |
| $\lvert-\rangle_{A_2}\lvert+\rangle_{A_2'}$ | $\lvert+\rangle_{B_2}\lvert+\rangle_{B_2'}$ | $\lvert+\rangle_{C_2}\lvert+\rangle_{C_2'}$ | $\hat{I}_{a_2}\otimes\hat{I}_{a_3}\otimes\hat{I}_{a_2'}\otimes\hat{I}_{a_3'}$ | $\hat{\sigma}^z_{b_2}\otimes\hat{I}_{b_3}\otimes\hat{I}_{b_2'}\otimes\hat{I}_{b_3'}$ | $\hat{I}_{c_2}\otimes\hat{I}_{c_3}\otimes\hat{I}_{c_2'}\otimes\hat{I}_{c_3'}$ |
| $\lvert-\rangle_{A_2}\lvert-\rangle_{A_2'}$ | $\lvert+\rangle_{B_2}\lvert+\rangle_{B_2'}$ | $\lvert+\rangle_{C_2}\lvert+\rangle_{C_2'}$ | $\hat{I}_{a_2}\otimes\hat{I}_{a_3}\otimes\hat{I}_{a_2'}\otimes\hat{I}_{a_3'}$ | $\hat{\sigma}^z_{b_2}\otimes\hat{I}_{b_3}\otimes\hat{I}_{b_2'}\otimes\hat{I}_{b_3'}$ | $\hat{I}_{c_2}\otimes\hat{I}_{c_3}\otimes\hat{I}_{c_2'}\otimes\hat{\sigma}^z_{c_3'}$ |
| $\lvert+\rangle_{A_2}\lvert+\rangle_{A_2'}$ | $\lvert+\rangle_{B_2}\lvert-\rangle_{B_2'}$ | $\lvert+\rangle_{C_2}\lvert+\rangle_{C_2'}$ | $\hat{I}_{a_2}\otimes\hat{I}_{a_3}\otimes\hat{I}_{a_2'}\otimes\hat{I}_{a_3'}$ | $\hat{I}_{b_2}\otimes\hat{I}_{b_3}\otimes\hat{I}_{b_2'}\otimes\hat{I}_{b_3'}$ | $\hat{\sigma}^z_{c_2}\otimes\hat{I}_{c_3}\otimes\hat{I}_{c_2'}\otimes\hat{I}_{c_3'}$ |
| $\lvert+\rangle_{A_2}\lvert-\rangle_{A_2'}$ | $\lvert+\rangle_{B_2}\lvert-\rangle_{B_2'}$ | $\lvert+\rangle_{C_2}\lvert+\rangle_{C_2'}$ | $\hat{I}_{a_2}\otimes\hat{I}_{a_3}\otimes\hat{I}_{a_2'}\otimes\hat{I}_{a_3'}$ | $\hat{I}_{b_2}\otimes\hat{I}_{b_3}\otimes\hat{I}_{b_2'}\otimes\hat{I}_{b_3'}$ | $\hat{\sigma}^z_{c_2}\otimes\hat{I}_{c_3}\otimes\hat{\sigma}^z_{c_2'}\otimes\hat{I}_{c_3'}$ |
| $\lvert-\rangle_{A_2}\lvert+\rangle_{A_2'}$ | $\lvert+\rangle_{B_2}\lvert-\rangle_{B_2'}$ | $\lvert+\rangle_{C_2}\lvert+\rangle_{C_2'}$ | $\hat{I}_{a_2}\otimes\hat{I}_{a_3}\otimes\hat{I}_{a_2'}\otimes\hat{I}_{a_3'}$ | $\hat{\sigma}^z_{b_3}\otimes\hat{I}_{b_3}\otimes\hat{I}_{b_2'}\otimes\hat{I}_{b_3'}$ | $\hat{\sigma}^z_{c_2}\otimes\hat{I}_{c_3}\otimes\hat{I}_{c_2'}\otimes\hat{I}_{c_3'}$ |
| $\lvert-\rangle_{A_2}\lvert-\rangle_{A_2'}$ | $\lvert+\rangle_{B_2}\lvert-\rangle_{B_2'}$ | $\lvert+\rangle_{C_2}\lvert+\rangle_{C_2'}$ | $\hat{I}_{a_2}\otimes\hat{I}_{a_3}\otimes\hat{I}_{a_2'}\otimes\hat{I}_{a_3'}$ | $\hat{\sigma}^z_{b_3}\otimes\hat{I}_{b_3}\otimes\hat{I}_{b_2'}\otimes\hat{I}_{b_3'}$ | $\hat{\sigma}^z_{c_2}\otimes\hat{I}_{c_3}\otimes\hat{I}_{c_2'}\otimes\hat{\sigma}^z_{c_3'}$ |
| $\lvert+\rangle_{A_2}\lvert+\rangle_{A_2'}$ | $\lvert-\rangle_{B_2}\lvert+\rangle_{B_2'}$ | $\lvert+\rangle_{C_2}\lvert+\rangle_{C_2'}$ | $\hat{\sigma}^z_{a_2}\otimes\hat{I}_{a_3}\otimes\hat{I}_{a_2'}\otimes\hat{I}_{a_3'}$ | $\hat{I}_{b_2}\otimes\hat{I}_{b_3}\otimes\hat{I}_{b_2'}\otimes\hat{I}_{b_3'}$ | $\hat{\sigma}^z_{c_2}\otimes\hat{I}_{c_3}\otimes\hat{I}_{c_2'}\otimes\hat{I}_{c_3'}$ |
| $\lvert+\rangle_{A_2}\lvert-\rangle_{A_2'}$ | $\lvert-\rangle_{B_2}\lvert+\rangle_{B_2'}$ | $\lvert+\rangle_{C_2}\lvert+\rangle_{C_2'}$ | $\hat{\sigma}^z_{a_2}\otimes\hat{I}_{a_3}\otimes\hat{I}_{a_2'}\otimes\hat{I}_{a_3'}$ | $\hat{I}_{b_2}\otimes\hat{I}_{b_3}\otimes\hat{I}_{b_2'}\otimes\hat{I}_{b_3'}$ | $\hat{\sigma}^z_{c_2}\otimes\hat{I}_{c_3}\otimes\hat{I}_{c_2'}\otimes\hat{\sigma}^z_{c_3'}$ |
| $\lvert-\rangle_{A_2}\lvert+\rangle_{A_2'}$ | $\lvert-\rangle_{B_2}\lvert+\rangle_{B_2'}$ | $\lvert+\rangle_{C_2}\lvert+\rangle_{C_2'}$ | $\hat{\sigma}^z_{a_2}\otimes\hat{I}_{a_3}\otimes\hat{I}_{a_2'}\otimes\hat{I}_{a_3'}$ | $\hat{\sigma}^z_{b_3}\otimes\hat{I}_{b_3}\otimes\hat{I}_{b_2'}\otimes\hat{I}_{b_3'}$ | $\hat{\sigma}^z_{c_2}\otimes\hat{I}_{c_3}\otimes\hat{I}_{c_2'}\otimes\hat{I}_{c_3'}$ |
| $\lvert-\rangle_{A_2}\lvert-\rangle_{A_2'}$ | $\lvert-\rangle_{B_2}\lvert+\rangle_{B_2'}$ | $\lvert+\rangle_{C_2}\lvert+\rangle_{C_2'}$ | $\hat{\sigma}^z_{a_2}\otimes\hat{I}_{a_3}\otimes\hat{I}_{a_2'}\otimes\hat{I}_{a_3'}$ | $\hat{\sigma}^z_{b_3}\otimes\hat{I}_{b_3}\otimes\hat{I}_{b_2'}\otimes\hat{I}_{b_3'}$ | $\hat{\sigma}^z_{c_2}\otimes\hat{I}_{c_3}\otimes\hat{I}_{c_2'}\otimes\hat{I}_{c_3'}$ |
| $\lvert+\rangle_{A_2}\lvert+\rangle_{A_2'}$ | $\lvert-\rangle_{B_2}\lvert-\rangle_{B_2'}$ | $\lvert+\rangle_{C_2}\lvert+\rangle_{C_2'}$ | $\hat{\sigma}^z_{a_2}\otimes\hat{I}_{a_3}\otimes\hat{I}_{a_2'}\otimes\hat{I}_{a_3'}$ | $\hat{I}_{b_2}\otimes\hat{I}_{b_3}\otimes\hat{I}_{b_2'}\otimes\hat{I}_{b_3'}$ | $\hat{\sigma}^z_{c_2}\otimes\hat{I}_{c_3}\otimes\hat{I}_{c_2'}\otimes\hat{I}_{c_3'}$ |
| $\lvert+\rangle_{A_2}\lvert-\rangle_{A_2'}$ | $\lvert-\rangle_{B_2}\lvert-\rangle_{B_2'}$ | $\lvert+\rangle_{C_2}\lvert+\rangle_{C_2'}$ | $\hat{\sigma}^z_{a_2}\otimes\hat{I}_{a_3}\otimes\hat{I}_{a_2'}\otimes\hat{I}_{a_3'}$ | $\hat{I}_{b_2}\otimes\hat{I}_{b_3}\otimes\hat{I}_{b_2'}\otimes\hat{I}_{b_3'}$ | $\hat{\sigma}^z_{c_2}\otimes\hat{I}_{c_3}\otimes\hat{I}_{c_2'}\otimes\hat{\sigma}^z_{c_3'}$ |
| $\lvert-\rangle_{A_2}\lvert+\rangle_{A_2'}$ | $\lvert-\rangle_{B_2}\lvert-\rangle_{B_2'}$ | $\lvert+\rangle_{C_2}\lvert+\rangle_{C_2'}$ | $\hat{\sigma}^z_{a_2}\otimes\hat{I}_{a_3}\otimes\hat{I}_{a_2'}\otimes\hat{I}_{a_3'}$ | $\hat{\sigma}^z_{b_3}\otimes\hat{I}_{b_3}\otimes\hat{I}_{b_2'}\otimes\hat{I}_{b_3'}$ | $\hat{\sigma}^z_{c_2}\otimes\hat{I}_{c_3}\otimes\hat{I}_{c_2'}\otimes\hat{I}_{c_3'}$ |
| $\lvert - \rangle_{A_2} \lvert - \rangle_{A'_2}$ | $\lvert - \rangle_{B_2} \lvert - \rangle_{B'_2}$ | $\lvert + \rangle_{C_2} \lvert + \rangle_{C'_2}$ | $\hat{\sigma}^z_{a_2} \otimes \hat{I}_{a_3} \otimes \hat{I}_{a'_2} \otimes \hat{I}_{a'_3}$ | $\hat{\sigma}^z_{b_3} \otimes \hat{I}_{b_3} \otimes \hat{I}_{b'_2} \otimes \hat{I}_{b'_3}$ | $\hat{\sigma}^z_{c_2} \otimes \hat{I}_{c_3} \otimes \hat{I}_{c'_2} \otimes \hat{\sigma}^z_{c'_3}$ |
| $\lvert + \rangle_{A_2} \lvert + \rangle_{A'_2}$ | $\lvert + \rangle_{B_2} \lvert + \rangle_{B'_2}$ | $\lvert + \rangle_{C_2} \lvert - \rangle_{C'_2}$ | $\hat{I}_{a_2} \otimes \hat{I}_{a_3} \otimes \hat{\sigma}^z_{a'_2} \otimes \hat{I}_{a'_3}$ | $\hat{I}_{b_2} \otimes \hat{I}_{b_3} \otimes \hat{I}_{b'_2} \otimes \hat{I}_{b'_3}$ | $\hat{I}_{c_2} \otimes \hat{I}_{c_3} \otimes \hat{I}_{c'_2} \otimes \hat{I}_{c'_3}$ |
| $\lvert + \rangle_{A_2} \lvert - \rangle_{A'_2}$ | $\lvert + \rangle_{B_2} \lvert + \rangle_{B'_2}$ | $\lvert + \rangle_{C_2} \lvert - \rangle_{C'_2}$ | $\hat{I}_{a_2} \otimes \hat{I}_{a_3} \otimes \hat{\sigma}^z_{a'_2} \otimes \hat{I}_{a'_3}$ | $\hat{I}_{b_2} \otimes \hat{I}_{b_3} \otimes \hat{I}_{b'_2} \otimes \hat{I}_{b'_3}$ | $\hat{I}_{c_2} \otimes \hat{I}_{c_3} \otimes \hat{I}_{c'_2} \otimes \hat{\sigma}^z_{c'_3}$ |
| $\lvert - \rangle_{A_2} \lvert + \rangle_{A'_2}$ | $\lvert + \rangle_{B_2} \lvert + \rangle_{B'_2}$ | $\lvert + \rangle_{C_2} \lvert - \rangle_{C'_2}$ | $\hat{\sigma}^z_{a_2} \otimes \hat{I}_{a_3} \otimes \hat{\sigma}^z_{a'_2} \otimes \hat{I}_{a'_3}$ | $\hat{I}_{b_2} \otimes \hat{I}_{b_3} \otimes \hat{\sigma}^z_{b'_2} \otimes \hat{I}_{b'_3}$ | $\hat{I}_{c_2} \otimes \hat{I}_{c_3} \otimes \hat{I}_{c'_2} \otimes \hat{I}_{c'_3}$ |
| $\lvert - \rangle_{A_2} \lvert - \rangle_{A'_2}$ | $\lvert + \rangle_{B_2} \lvert + \rangle_{B'_2}$ | $\lvert + \rangle_{C_2} \lvert - \rangle_{C'_2}$ | $\hat{I}_{a_2} \otimes \hat{I}_{a_3} \otimes \hat{\sigma}^z_{a'_2} \otimes \hat{I}_{a'_3}$ | $\hat{\sigma}^z_{b_2} \otimes \hat{I}_{b_3} \otimes \hat{I}_{b'_2} \otimes \hat{I}_{b'_3}$ | $\hat{I}_{c_2} \otimes \hat{I}_{c_3} \otimes \hat{I}_{c'_2} \otimes \hat{\sigma}^z_{c'_3}$ |
| $\lvert + \rangle_{A_2} \lvert + \rangle_{A'_2}$ | $\lvert + \rangle_{B_2} \lvert - \rangle_{B'_2}$ | $\lvert + \rangle_{C_2} \lvert - \rangle_{C'_2}$ | $\hat{I}_{a_2} \otimes \hat{I}_{a_3} \otimes \hat{\sigma}^z_{a'_2} \otimes \hat{I}_{a'_3}$ | $\hat{I}_{b_2} \otimes \hat{I}_{b_3} \otimes \hat{I}_{b'_2} \otimes \hat{I}_{b'_3}$ | $\hat{\sigma}^z_{c_2} \otimes \hat{I}_{c_3} \otimes \hat{I}_{c'_2} \otimes \hat{I}_{c'_3}$ |
| $\lvert + \rangle_{A_2} \lvert - \rangle_{A'_2}$ | $\lvert + \rangle_{B_2} \lvert - \rangle_{B'_2}$ | $\lvert + \rangle_{C_2} \lvert - \rangle_{C'_2}$ | $\hat{I}_{a_2} \otimes \hat{I}_{a_3} \otimes \hat{\sigma}^z_{a'_2} \otimes \hat{I}_{a'_3}$ | $\hat{I}_{b_2} \otimes \hat{I}_{b_3} \otimes \hat{I}_{b'_2} \otimes \hat{I}_{b'_3}$ | $\hat{\sigma}^z_{c_2} \otimes \hat{I}_{c_3} \otimes \hat{I}_{c'_2} \otimes \hat{\sigma}^z_{c'_3}$ |
| $\lvert - \rangle_{A_2} \lvert + \rangle_{A'_2}$ | $\lvert + \rangle_{B_2} \lvert - \rangle_{B'_2}$ | $\lvert + \rangle_{C_2} \lvert - \rangle_{C'_2}$ | $\hat{I}_{a_2} \otimes \hat{I}_{a_3} \otimes \hat{\sigma}^z_{a'_2} \otimes \hat{I}_{a'_3}$ | $\hat{\sigma}^z_{b_2} \otimes \hat{I}_{b_3} \otimes \hat{I}_{b'_2} \otimes \hat{I}_{b'_3}$ | $\hat{\sigma}^z_{c_2} \otimes \hat{I}_{c_3} \otimes \hat{I}_{c'_2} \otimes \hat{I}_{c'_3}$ |
| $\lvert - \rangle_{A_2} \lvert - \rangle_{A'_2}$ | $\lvert + \rangle_{B_2} \lvert - \rangle_{B'_2}$ | $\lvert + \rangle_{C_2} \lvert - \rangle_{C'_2}$ | $\hat{I}_{a_2} \otimes \hat{I}_{a_3} \otimes \hat{\sigma}^z_{a'_2} \otimes \hat{I}_{a'_3}$ | $\hat{\sigma}^z_{b_2} \otimes \hat{I}_{b_3} \otimes \hat{I}_{b'_2} \otimes \hat{I}_{b'_3}$ | $\hat{\sigma}^z_{c_2} \otimes \hat{I}_{c_3} \otimes \hat{I}_{c'_2} \otimes \hat{\sigma}^z_{c'_3}$ |
| $\lvert + \rangle_{A_2} \lvert + \rangle_{A'_2}$ | $\lvert - \rangle_{B_2} \lvert + \rangle_{B'_2}$ | $\lvert + \rangle_{C_2} \lvert - \rangle_{C'_2}$ | $\hat{\sigma}^z_{a_2} \otimes \hat{I}_{a_3} \otimes \hat{\sigma}^z_{a'_2} \otimes \hat{I}_{a'_3}$ | $\hat{I}_{b_2} \otimes \hat{I}_{b_3} \otimes \hat{I}_{b'_2} \otimes \hat{I}_{b'_3}$ | $\hat{I}_{c_2} \otimes \hat{I}_{c_3} \otimes \hat{I}_{c'_2} \otimes \hat{I}_{c'_3}$ |
| $\lvert + \rangle_{A_2} \lvert - \rangle_{A'_2}$ | $\lvert - \rangle_{B_2} \lvert + \rangle_{B'_2}$ | $\lvert + \rangle_{C_2} \lvert - \rangle_{C'_2}$ | $\hat{\sigma}^z_{a_2} \otimes \hat{I}_{a_3} \otimes \hat{\sigma}^z_{a'_2} \otimes \hat{I}_{a'_3}$ | $\hat{I}_{b_2} \otimes \hat{I}_{b_3} \otimes \hat{I}_{b'_2} \otimes \hat{I}_{b'_3}$ | $\hat{I}_{c_2} \otimes \hat{I}_{c_3} \otimes \hat{I}_{c'_2} \otimes \hat{\sigma}^z_{c'_3}$ |
| $\lvert - \rangle_{A_2} \lvert + \rangle_{A'_2}$ | $\lvert - \rangle_{B_2} \lvert + \rangle_{B'_2}$ | $\lvert + \rangle_{C_2} \lvert - \rangle_{C'_2}$ | $\hat{\sigma}^z_{a_2} \otimes \hat{I}_{a_3} \otimes \hat{\sigma}^z_{a'_2} \otimes \hat{I}_{a'_3}$ | $\hat{\sigma}^z_{a_2} \otimes \hat{I}_{a_3} \otimes \hat{\sigma}^z_{a'_2} \otimes \hat{I}_{a'_3}$ | $\hat{I}_{c_2} \otimes \hat{I}_{c_3} \otimes \hat{I}_{c'_2} \otimes \hat{I}_{c'_3}$ |
| $\lvert - \rangle_{A_2} \lvert - \rangle_{A'_2}$ | $\lvert - \rangle_{B_2} \lvert + \rangle_{B'_2}$ | $\lvert + \rangle_{C_2} \lvert - \rangle_{C'_2}$ | $\hat{\sigma}^z_{a_2} \otimes \hat{I}_{a_3} \otimes \hat{\sigma}^z_{a'_2} \otimes \hat{I}_{a'_3}$ | $\hat{\sigma}^z_{b_2} \otimes \hat{I}_{b_3} \otimes \hat{I}_{b'_2} \otimes \hat{I}_{b'_3}$ | $\hat{I}_{c_2} \otimes \hat{I}_{c_3} \otimes \hat{I}_{c'_2} \otimes \hat{\sigma}^z_{c'_3}$ |
| $\lvert + \rangle_{A_2} \lvert + \rangle_{A'_2}$ | $\lvert - \rangle_{B_2} \lvert - \rangle_{B'_2}$ | $\lvert + \rangle_{C_2} \lvert - \rangle_{C'_2}$ | $\hat{\sigma}^z_{a_2} \otimes \hat{I}_{a_3} \otimes \hat{\sigma}^z_{a'_2} \otimes \hat{I}_{a'_3}$ | $\hat{I}_{b_2} \otimes \hat{I}_{b_3} \otimes \hat{I}_{b'_2} \otimes \hat{I}_{b'_3}$ | $\hat{\sigma}^z_{c_2} \otimes \hat{I}_{c_3} \otimes \hat{I}_{c'_2} \otimes \hat{I}_{c'_3}$ |
| $\lvert + \rangle_{A_2} \lvert - \rangle_{A'_2}$ | $\lvert - \rangle_{B_2} \lvert - \rangle_{B'_2}$ | $\lvert + \rangle_{C_2} \lvert - \rangle_{C'_2}$ | $\hat{\sigma}^z_{a_2} \otimes \hat{I}_{a_3} \otimes \hat{\sigma}^z_{a'_2} \otimes \hat{I}_{a'_3}$ | $\hat{I}_{b_2} \otimes \hat{I}_{b_3} \otimes \hat{I}_{b'_2} \otimes \hat{I}_{b'_3}$ | $\hat{\sigma}^z_{c_2} \otimes \hat{I}_{c_3} \otimes \hat{I}_{c'_2} \otimes \hat{\sigma}^z_{c'_3}$ |
| $\lvert - \rangle_{A_2} \lvert + \rangle_{A'_2}$ | $\lvert - \rangle_{B_2} \lvert - \rangle_{B'_2}$ | $\lvert + \rangle_{C_2} \lvert - \rangle_{C'_2}$ | $\hat{\sigma}^z_{a_2} \otimes \hat{I}_{a_3} \otimes \hat{\sigma}^z_{a'_2} \otimes \hat{I}_{a'_3}$ | $\hat{\sigma}^z_{b_2} \otimes \hat{I}_{b_3} \otimes \hat{I}_{b'_2} \otimes \hat{I}_{b'_3}$ | $\hat{\sigma}^z_{c_2} \otimes \hat{I}_{c_3} \otimes \hat{I}_{c'_2} \otimes \hat{I}_{c'_3}$ |
| $\lvert - \rangle_{A_2} \lvert - \rangle_{A'_2}$ | $\lvert - \rangle_{B_2} \lvert - \rangle_{B'_2}$ | $\lvert + \rangle_{C_2} \lvert - \rangle_{C'_2}$ | $\hat{\sigma}^z_{a_2} \otimes \hat{I}_{a_3} \otimes \hat{\sigma}^z_{a'_2} \otimes \hat{I}_{a'_3}$ | $\hat{\sigma}^z_{b_2} \otimes \hat{I}_{b_3} \otimes \hat{I}_{b'_2} \otimes \hat{I}_{b'_3}$ | $\hat{\sigma}^z_{c_2} \otimes \hat{I}_{c_3} \otimes \hat{I}_{c'_2} \otimes \hat{\sigma}^z_{c'_3}$ |